# Stochastic theory of ferroelectric domain structure formation dominated by quenched disorder

Olga Y. Mazur[1,2], Leonid I. Stefanovich[1] and Yuri A. Genenko[2]

[1]Branch for Physics of Mining Processes of the M.S. Poliakov Institute of Geotechnical Mechanics of the National Academy of Sciences of Ukraine, 49600 Dnipro, Simferopolska st., 2a, Ukraine

[2]Institute of Materials Science, Technical University of Darmstadt, Otto-Berndt-Str, 3, 64287 Darmstadt, Germany

A self-consistent stochastic model of domain structure formation in a uniaxial ferroelectric, quenched from a high-temperature paraelectric phase to a low-temperature ferroelectric phase, is developed with an account of the applied electric field and the feedback effect via local depolarization fields. Both polarization and field components are considered as Gauss random variables. A system of integro-differential equations for correlation functions of all involved variables is derived and solved analytically and numerically. Phase diagram in terms of the average value and dispersion of polarization reveals different possible equilibrium states and available final single-domain and multi-domain states. The time-dependent evolution of the average polarization and dispersion discloses a bifurcation behavior and the temperature-dependent value of the electric field, deciding between the single-domain and multi-domain final states, which can be interpreted as the coercive field. Analytical and numerical results for the time-dependent correlation length and correlation functions exhibit plausible agreement with available experimental data.

## 1. Introduction

Domain structures in ferroelectrics have a decisive effect on their fundamental physical properties and functionality, so that the domain engineering is a prerequisite for many ferroelectric implementations like piezoelectric sensor and actuator technology [1], nonlinear optics and photonics [2]. A possible way of formation of domain structures with desired characteristics is the controlled quenching from a high-temperature paraelectric to a low-temperature ferroelectric phase. This intrinsically stochastic process depending on material properties, sample shape, quenching temperature, and applied field regimes was a subject of experimental studies over many decades. One of the most intensively investigated systems was thereby single-crystalline triglycine sulfate (TGS) because this model ferroelectric material possesses the only polar axis and exhibits only 180°-domains [3-17]. A thorough review of experimental methods used for visualization of static and dynamic domain structures in TGS including the etching and decoration methods, the powder deposition technique, the nematic liquid crystal coating, the scanning electron microscopy, the X-ray diffraction microscopy, and others was given by Nakatani [4]. Later an in-situ 3D observation of domain formation dynamics near a ferroelectric–paraelectric phase transition was performed by Wehmeier et al. [14] using the second harmonic generation microscopy.

Temperature dependence of domain patterns in TGS, represented by isolated lenticular domains and extended lamellar domains, was investigated in Refs. [9,10-16] in different annealing and cooling regimes. The evolution of domain structures with time after quenching at different temperatures was studied in Refs. [3,5-9,14,16]. Main features of time-dependent characteristics, represented essentially by the correlation length and the two-site correlation function, were revealed already in the pioneering works by Nakatani [3] and the Ishibashi group [5,6]. A dominating characteristic length $L(t)$ was found to grow with time $t$ as the power function $L(t) \sim (t - t_0)^\nu$, where the exponent $\nu$ was reported to adopt values about 0.2-0.3 [5,6], 1/3 [7,8], 0.086-0.155 [9], 0.3 [10] while the offset time $t_0$ was either positive [8,9] or negative [5-7,9], revealing then a finite initial value $L(0)$. Golitsyna et al. reported that $\nu$ takes values about 0.3 at temperature $T$ difference from the transition temperature $T_c - T > 1\,K$ but rapidly approaches unity when $T_c - T < 0.3\,K$ [16] confirming the observation by Nakatani [3]. At longer times the asymptotic dependence $L(t) \sim [\ln(t/t_0)]^4$ was observed [7-9]. Tolstikhina et al. applied the Fourier analysis to the 2D domain pattern allowing a consistent quantification of the correlation length [13].

A time-dependent two-site correlation function was introduced as $\langle \pi(\mathbf{r}_1, t) \pi(\mathbf{r}_2, t) \rangle$, where $\mathbf{s} = \mathbf{r}_1 - \mathbf{r}_2$ and $\pi(\mathbf{r}, t) = P_z(\mathbf{r}, t)/P_s$ is the normalized polarization component with the spontaneous polarization $P_s$ [5,6]. The statistical average $\langle ... \rangle$ was determined experimentally as the spatial average over the sample volume. The correlation function turned out to be anisotropic in the plane perpendicular to the polarization direction $z$ and demonstrated oscillations in the direction perpendicular to the typical lamellar domain alignment, while no oscillations were observed along the lamellas [5,16]. Scaling features of the correlation function represented as a function of $s/L(t)$ were established at least at short distances $s$ [6,7-9,16]. For the system with applied total [5] or local [17] electric fields the dynamical scaling features were observed too.

Initial theoretical considerations of the stochastic domain formation were mostly based on either the time-dependent Landau-Ginzburg-Devonshire (LGD) model or the Kolmogorov-Avrami-Ishibashi (KAI) statistical approach. Thus, a stochastic formation of domain structures in a ferroelectric subject to a periodic electric field was studied in a KAI-based model neglecting statistical correlations between nucleating domains [18]. Together with the empirical knowledge on the electrical field dependence of the domain wall velocity and relaxation time [19] the KAI approach allowed an explanation of the time dependence of the field-driven switching of the average polarization [20], however the correlation radius and correlation function behavior remained elusive.

The first attempt of the LGD-based stochastic analysis of the domain formation was made by Rao and Chakrabarti [21]. In this seminal work, the frozen-in disorder (quenched random field) as well as the thermodynamic fluctuations represented by the white noise were accounted. Numerical integration of the time-dependent LGD equation for a one-component order parameter revealed an initial diffusive regime of the correlation length with $L(t) \sim t^{1/2}$ followed by an asymptotic logarithmic dependence due to the frozen-in randomness. Correlation properties of the domain growth were not studied.

Darinskii et al. considered the LGD model with a two-component polarization order parameter accounting self-consistently for the electric depolarization fields emerging from the polarization inhomogeneities [22]. Deterministic solutions of the coupled LGD and Poisson equations revealed single-domain and periodic structures appearing in different regimes. The spatial period was derived from the condition of the maximum growth rate of the order parameter and could be related to the characteristic length $L(t)$, however, the transient development of the inhomogeneous phase was not studied.

The phase-field approach to ferroelectrics generalizing the LGD theory considers typically much more complicated problem statements including multi-component polarization and elastic variables, their electro-elastic [23] and flexoelectric [24,25] coupling, together with other degrees of freedom and features, such as long-range dipole-dipole interactions [26], semiconducting effects and mobile charged defects [27,28]. Analytical treatment of such problems is usually not possible, but a great advantage of phase-field simulations is that they allow simultaneous accounting of multiple physical effects resulting in structures and features which could hardly be conceived theoretically [23-30]. New perspectives were opened by molecular dynamics simulations of lead titanate, based on interatomic potentials parameterized from first-principles, supported by the phase-field simulations with properly adjusted parameters [31,32]. Particularly, molecular dynamics gave insight in microscopic mechanisms of the field-driven domain switching in perovskites [31-33].

An LGD-based stochastic treatment of the domain formation in the presence of an external weak electric field in uniaxial ferroelectrics dominated by the quenched disorder was elaborated by Stefanovich in terms of the time-dependent two-site correlation function and the mean polarization [34]. It was analytically shown that the correlation length grows with time as $L(t) = \sqrt{L^2(0) + 2t/3}$ starting with an initial value $L(0)$. A phase diagram in coordinates of the polarization dispersion and the average polarization was constructed which demonstrated a tendency to the formation of single-domain states at lower temperatures and multi-domain states at higher temperatures closer to the transition point. The further numerical study of this model disclosed generally non-monotonic time-dependences of both the mean polarization and the polarization dispersion revealing a characteristic applied field deciding between the single-domain and multi-domain asymptotic state [35]. The approach proved to be fruitful when applied to the problem of domain formation under an applied ac field [36] as well as by the construction of pressure-temperature diagram of phase states of a barium titanate single-crystal confirmed experimentally [37]. The drawback of the approach [34-37] is an assumption of a uniform electric field in the ferroelectric hardly compatible with the formation of stochastic nonuniform polarization domain structures.

In the current study, we extend the LGD-based stochastic approach suggested in Ref. [34] by introducing stochastic electric field variables, self-consistently related to the polarization via the Poisson equation. We derive a system of integro-differential equations for self- and cross-correlation functions for all involved stochastic variables considering them as Gauss random variables and solve these equations analytically and numerically. The results for the time-dependent correlation length and correlation functions are compared with available experiments [3,5-9,15,16]. In spite of the superficial similarity with the previous considerations [34-37] the actual approach is physically quite different. Random spatial variations of polarization in the studied system create local charge density, which in turn generates substantial stochastic depolarization fields. These fields have a great impact on the domain formation and development with time. In the previous studies [34-37], the electric field in the system was assumed to be equal to a uniform external field and the appearance of the depolarization fields was neglected, therefore the consideration was, in fact, limited to infinitesimal external fields. In the current work, the electric field includes both the external one and the emerging depolarization fields described self-consistently via the Poisson equation with proper boundary conditions. This allows consideration of a finite sample subject to arbitrary applied electric fields and a construction of the parametric phase diagram. The final state of the evolution of the system, which can be a single- or multi-domain one, is determined by the external applied electric field, and the quenching temperature. The characteristics of this state given by the mean polarization and the polarization dispersion (variance) determine the functional properties of the ferroelectric. Among others we show a crucial impact of cross-correlations between the electric field and polarization on the choice between a single- and a multi-domain final state.

## 2. LGD-based stochastic model
### A. LGD model of a uniaxial ferroelectric/nonferroelastic

We consider a uniaxial single-crystalline ferroelectric exhibiting only 180° polarization domains as in the case of TGS, LiNbO$_3$, and LiTaO$_3$ materials. We study the evolution of the system from a disordered initial state obtained by quenching from the high-temperature paraelectric phase to the ferroelectric phase at some fixed temperature $T < T_c$. In this model, we assume the stochastic formation of domain structures dominated by initial quenched polarization disorder. The conditions at which thermodynamic fluctuations can be neglected in comparison with the quenched disorder are derived in Appendix A and apply in the temperature region $T_c - T > 0.02\,K$ if the characteristic length of the initial disorder $L(0)$ exceeds the length scale of thermodynamic fluctuations.

We choose a typical experimental geometry [3-6,14] of a single-crystalline ferroelectric plate of thickness $h_f$, attached to a bottom metallic electrode and a dielectric layer of thickness $h_d$ at the top side (see Fig. 1) covered with a top metallic electrode. Both electrodes are kept at fixed electric potentials providing a desired applied electric field regime. The Gibbs free energy of the system can be presented in the form [23,38]

$$\Phi = \Phi_0 + \int_{V_f} \left[ \frac{1}{2} A P_z^2 + \frac{1}{4} B P_z^4 + \frac{1}{2} G (\nabla P_z)^2 - P_z E_z - \frac{\varepsilon_0 \varepsilon_b}{2} \boldsymbol{E}^2 \right] dV - \int_{V_d} \frac{\varepsilon_0 \varepsilon_d}{2} \boldsymbol{E}^2 dV \qquad (1)$$

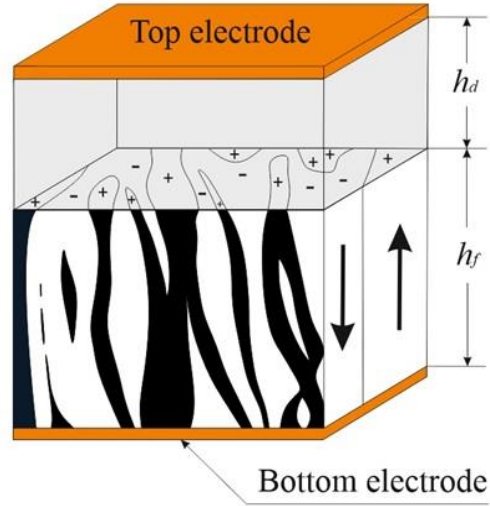

Fig. 1. Problem layout: A ferroelectric slab of thickness $h_f$, attached to a bottom electrode and separated from a top electrode by a dielectric layer of thickness $h_d$, is infinite in a plane parallel to the slab surface. Polarization direction is along the vertical $z$-axis of the Cartesian $(x, y, z)$-frame. The scheme of 180°-domains roughly reproduces the domain pattern observed in Ref. [15].

where $A = \alpha_0 (T - T_c)$, $\alpha_0 > 0$, $T < T_c$, which is the temperature of the paraelectric-ferroelectric phase transition. The other coefficients of the LGD expansion are $B > 0$ and $G > 0$. This form accounts for the only polarization component $P_z$ in the Cartesian frame $(x, y, z)$ perpendicular to the ferroelectric plate surface and allows description of the second-order phase transition which is the case for TGS. Polarization is the primary order parameter of the phase transition; elastic variables are the secondary order parameters and can be neglected in TGS since it is nonferroelastic. $\boldsymbol{E}$ is the local electric field, $\varepsilon_0$, $\varepsilon_d$ and $\varepsilon_b$ are the permittivity of vacuum, of the dielectric layer and the background permittivity of the ferroelectric, respectively. $E_z$ denotes the z-component of the local electric field which may originate from external sources and/or from spatial variation of polarization. $V_f$ and $V_d$ denote the volumes of the ferroelectric plate and the dielectric layer, respectively.

The according Landau-Khalatnikov equation governing the evolution of polarization reads

$$\Gamma \frac{\partial P_z}{\partial t} = -\frac{\delta \Phi}{\delta P_z} = -AP_z - BP_z^3 + G\Delta P_z + E_z, \qquad (2)$$

with the Khalatnikov constant $\Gamma$ and the Laplace operator $\Delta$. Variation of the electric field $\boldsymbol{E} = -\nabla \varphi$, where $\varphi$ is the electric potential, is described by the Poisson equation in the ferroelectric,

$$\varepsilon_0 \varepsilon_b \Delta \varphi = \frac{\partial P_z}{\partial z} \qquad (3)$$

and by the Laplace equation in the dielectric,

$$\Delta \varphi = 0, \qquad (4)$$

which can be derived by minimization of the functional (1) with respect to $\varphi$. The boundary conditions for the above equations include the continuity of the electric potential at all interfaces and of the normal electric displacement at the ferroelectric/dielectric interface [38],

$$\varphi|_{z=0} = 0, \ \varphi|_{z=h_f-0} = \varphi|_{z=h_f+0}, \ \varphi|_{z=h_f+h_d} = -V, \ D_z|_{z=h_f-0} = D_z|_{z=h_f+0}, \qquad (5)$$

where the electric displacement equals $D_z = \varepsilon_0 \varepsilon_b E_z + P_z$ in the ferroelectric and $D_z = \varepsilon_0 \varepsilon_d E_z$ in the dielectric.

B. Stochastic variables and their mean values

The evolution of the system from the initial disordered quenched state makes all involved physical variables time-dependent random variables. Statistical averaging of random variables is indicated with a symbol $\langle ... \rangle$ and is assumed to coincide with the average over the material volume. Thus, we introduce a local polarization $P_z(\boldsymbol{r}, t) = \langle P_z \rangle + \delta P_z(\boldsymbol{r}, t)$, where the mean polarization may be a function of time while $\langle \delta P_z \rangle = 0$. The mean values of the electric field in the ferroelectric and the dielectric regions are denoted $E_f$ and $E_d$, respectively. If the mean polarization value $\langle P_z \rangle$ is not zero, this creates a mean electric depolarization field in the system. The electric potential can be split in a regular and a stochastic part as $\varphi = \bar{\varphi} + \delta \varphi$, where $\langle \delta \varphi \rangle = 0$ and the regular part $\bar{\varphi}$ satisfies the continuity conditions (5). Thus,

$$\bar{\varphi}(z) = \begin{cases} -E_f z, & 0 < z < h_f \\ -E_f h_f - E_d(z - h_f), & h_f < z < h_f + h_d \end{cases}. \qquad (6)$$

By satisfying the boundary condition at $z = h_f + h_d$ and the continuity of the electric displacement (5) at $z = h_f$ one finds the expressions for the mean electric fields in both media, depending on the mean polarization,

$$E_d = \frac{\varepsilon_b}{\varepsilon_d h_f + \varepsilon_b h_d} V + \frac{h_f}{\varepsilon_d h_f + \varepsilon_b h_d} \frac{\langle P_z \rangle}{\varepsilon_0}, \quad E_f = E_a - \rho_z \frac{\langle P_z \rangle}{\varepsilon_0} \tag{7}$$

with

$$E_a = \frac{\varepsilon_d}{\varepsilon_d h_f + \varepsilon_b h_d} V, \quad \rho_z = \frac{h_d}{\varepsilon_d h_f + \varepsilon_b h_d}. \tag{8}$$

Note that the field is present in the whole structure even if the electrodes are short-circuited, $V = 0$. The local electric field in the ferroelectric is given by

$$\boldsymbol{E}(\mathbf{r}, t) = \boldsymbol{E}_a - \frac{\rho_z}{\varepsilon_0}\langle \boldsymbol{P} \rangle - \nabla \delta\varphi(\mathbf{r}, t) \tag{9}$$

with $\boldsymbol{E}_a = (0,0, E_a)$ and $\boldsymbol{P} = (0,0, P_z)$.

### C. Equations in dimensionless units

It is convenient to normalize physical variables to their natural characteristic magnitudes in the phase transition problem. This leads to a dimensionless polarization $\pi = P_z/P_s$ normalized to the spontaneous equilibrium polarization $P_s = \sqrt{|A|/B}$, and a dimensionless electric field $\boldsymbol{\epsilon} = \boldsymbol{E}/E_0$ with the value of $E_0 = P_s|A|$ close to the thermodynamic coercive field, $E_{cr} = 2P_s|A|/3\sqrt{3}$ [39]. Spatial coordinates are normalized to a characteristic length $\lambda = \sqrt{G/|A|}$ (the characteristic domain wall thickness) and time $t$ to a characteristic time $t_0 = \Gamma/|A|$, $\tau = t/t_0$. Thus, we introduce a dimensionless local electric field as

$$\boldsymbol{\epsilon}(\mathbf{r},\tau) = \boldsymbol{\epsilon}_a - \alpha_z \bar{\pi}(\tau)\hat{\boldsymbol{z}} - \nabla\phi(\mathbf{r},\tau) \tag{10}$$

where the first two terms represent the mean electric field in the ferroelectric along the z-axis, as in Eq. (9), with the dimensionless mean polarization magnitude $\bar{\pi}(\tau)=\langle P_z \rangle/P_s$, dimensionless field $\epsilon_a = E_a/E_0$, dimensionless stochastic potential $\phi = \delta\varphi/(E_0\lambda)$ and the coefficient $\alpha_z = \rho_z/(\varepsilon_0|A|)$, while the mean fluctuation depolarization field due to spatial polarization variations vanishes, $\langle -\nabla\phi(\mathbf{r},\tau) \rangle = 0$. The local dimensionless polarization along the z-axis can be represented as $\pi(\mathbf{r},\tau) = \bar{\pi}(\tau) + \xi(\mathbf{r},\tau)$ where $\xi = \delta P_z/P_s$ and $\langle \xi(\mathbf{r},\tau) \rangle = 0$.

Introducing the dimensionless variables reduces the governing equation (2) to

$$\frac{\partial \pi}{\partial \tau} = \Delta \pi + \pi - \pi^3 + \epsilon_z. \tag{11}$$

and the Poisson equation (3) to

$$\Delta \phi = \eta \frac{\partial \pi}{\partial Z} \tag{12}$$

with $Z = z/\lambda$ and a dimensionless parameter $\eta = 1/(\varepsilon_0 \varepsilon_b |A|)$. Considering the ferroelectric susceptibility below $T_c$ given by $\chi_f = 1/(2\varepsilon_0|A|)$ [39] reveals that the parameter $\eta = 2\chi_f/\varepsilon_b$ is virtually the ferroelectric susceptibility normalized to the background one.

## 3. Correlation functions and their governing equations

Assuming that all physical variables are Gaussian random fields [40], their correlation properties can be completely characterized by two-site autocorrelation functions for polarization, $K(\mathbf{s},\tau) = \langle \xi(\mathbf{r}_1,\tau)\xi(\mathbf{r}_2,\tau)\rangle$, and electric potential, $g(\mathbf{s},\tau) = \langle \phi(\mathbf{r}_1,\tau)\phi(\mathbf{r}_2,\tau)\rangle$, with $\mathbf{s} = \mathbf{r}_1 - \mathbf{r}_2$, together with cross-correlation functions $\Psi_{xz}(\mathbf{s},\tau) = \langle \epsilon_x(\mathbf{r}_1,\tau)\xi(\mathbf{r}_2,\tau)\rangle$, $\Psi_{yz}(\mathbf{s},\tau) = \langle \epsilon_y(\mathbf{r}_1,\tau)\xi(\mathbf{r}_2,\tau)\rangle$, $\Psi_{zz}(\mathbf{s},\tau) = \langle \epsilon_z(\mathbf{r}_1,\tau)\xi(\mathbf{r}_2,\tau)\rangle$. The two-site correlation functions depend only on the position difference $\mathbf{s}$ in infinite uniform media that is assumed to apply also in macroscopic finite samples. Correlation functions are a useful tool for describing the domain structure in ferroelectrics. For example, the function $K(\mathbf{s},\tau)$ determines the degree of similarity of different fragments of a static domain picture separated by a position vector $\mathbf{s}$ at a given time. The function $K(\mathbf{s},\tau)$ is at maximum when the fragments are completely identical, as expected within one domain, and 0 if there are no correlations between the fragments. Changing the sign of the function $K(\mathbf{s},\tau)$ indicates that it has entered the domain region with the opposite direction of polarization. Thus, it contains information on the characteristic size of the domains, the regularity of domain structure, and its development with time. The cross-correlation function $\Psi_{\alpha\beta}(\mathbf{s},\tau)$ reveals how sensitive the local polarization to the electric fields generated by polarization variation at the other locations is. Correlation functions were experimentally studied over thirty years [3,5-9,16] but their theoretical description is still missing. A couple of examples of correlation functions evaluated for regular domain structures are presented in Appendix B.

Considering the potential nature of electric field (10) the correlation function for the electric field components can be expressed via $g(\mathbf{s},\tau)$ as

$$\langle \epsilon_\alpha(\mathbf{r}_1,\tau)\epsilon_\beta(\mathbf{r}_2,\tau)\rangle = [\epsilon_\alpha - \alpha_z\bar{\pi}(\tau)]^2 \delta_{\alpha z}\delta_{\beta z} - \partial^2 g(\mathbf{s},\tau)/\partial s_\alpha \partial s_\beta \quad (13)$$

with indices $\alpha$ and $\beta$ adopting values $x$, $y$ and $z$. Cross-correlation functions $\Psi_{\alpha\beta}(\mathbf{s},\tau)$ and the autocorrelation function for the potential $g(\mathbf{s},\tau)$ can be related to each other when considering the Gauss equation according to Eq. (12),

$$\frac{\partial \epsilon_x(\mathbf{r}_2,\tau)}{\partial X_2} + \frac{\partial \epsilon_y(\mathbf{r}_2,\tau)}{\partial Y_2} + \frac{\partial \epsilon_z(\mathbf{r}_2,\tau)}{\partial Z_2} = -\eta \frac{\partial \xi(\mathbf{r}_2,\tau)}{\partial Z_2}. \quad (14)$$

By multiplying it with $\epsilon_x(\mathbf{r}_1,\tau)$ and consequent averaging and using Eq. (13) one obtains a relation

$$\frac{\partial}{\partial s_x}\Delta g(\mathbf{s},\tau) = \eta \frac{\partial}{\partial s_z}\Psi_{xz}(\mathbf{s},\tau). \quad (15)$$

Similarly, by multiplying Eq. (14) with $\epsilon_y(\mathbf{r}_1,\tau)$ and consequent averaging and using Eq. (13) one gets a relation

$$\frac{\partial}{\partial s_y}\Delta g(\mathbf{s},\tau)=\eta\frac{\partial}{\partial s_z}\Psi_{yz}(\mathbf{s},\tau). \tag{16}$$

In the same manner, by multiplying Eq. (14) with $\epsilon_z(\mathbf{r_1},\tau)$ and consequent averaging and using Eq. (13) one finds a relation

$$\frac{\partial}{\partial s_z}\Delta g(\mathbf{s},\tau)=\eta\frac{\partial}{\partial s_z}\Psi_{zz}(\mathbf{s},\tau). \tag{17}$$

The last equation means that $\Delta g(\mathbf{s},\tau)$ and $\eta\Psi_{zz}(\mathbf{s},\tau)$ can differ only by a constant. Since asymptotically at $s\to\infty$ the correlations must vanish, this constant equals zero and thus

$$\Delta g(\mathbf{s},\tau)=\eta\Psi_{zz}(\mathbf{s},\tau). \tag{18}$$

If in the above derivation $\mathbf{r}_1$ and $\mathbf{r}_2$ were interchanged the same formulas would result with $\Psi_{\alpha\beta}(-\mathbf{s},\tau)$ meaning that $\Psi_{\alpha\beta}(-\mathbf{s},\tau)=\Psi_{\alpha\beta}(\mathbf{s},\tau)$.

Similarly, a relation between $g(\mathbf{s},\tau)$ and $K(\mathbf{s},\tau)$ can be derived. To this end Eq. (14) taken at $(\mathbf{r}_1,\tau)$ is multiplied with $\xi(\mathbf{r}_2,\tau)$ and averaged. This results in equation

$$\frac{\partial\Psi_{xz}(\mathbf{s},\tau)}{\partial s_x}+\frac{\partial\Psi_{yz}(\mathbf{s},\tau)}{\partial s_y}+\frac{\partial\Psi_{zz}(\mathbf{s},\tau)}{\partial s_z}=-\eta\frac{\partial K(\mathbf{s},\tau)}{\partial s_z}. \tag{19}$$

By differentiating this relation with respect to $s_z$ and utilizing relations (15-17) one finds finally

$$\Delta^2 g(\mathbf{s},\tau)=-\eta^2\frac{\partial^2}{\partial s_z^2}K(\mathbf{s},\tau), \tag{20}$$

or, using Eq. (18),

$$\Delta\Psi_{zz}(\mathbf{s},\tau)=-\eta\frac{\partial^2}{\partial s_z^2}K(\mathbf{s},\tau). \tag{21}$$

We proceed now with the derivation of equations of evolution for the mean polarization $\bar{\pi}(\tau)$ and the correlation function $K(\mathbf{s},\tau)$. The first one can be derived by statistical averaging of Eq. (11). By doing this, we substitute $\pi(\mathbf{r},\tau)=\bar{\pi}(\tau)+\xi(\mathbf{r},\tau)$ into Eq. (11) and average the latter, taking into account that $\langle\Delta\pi\rangle=0$ and that the averages of odd number of Gaussian variables vanish, if they have zero central moments [40], which is the case for $\xi(\mathbf{r},\tau)$, so that $\langle\xi^3(\mathbf{r},\tau)\rangle=0$. This results in equation

$$\frac{d\bar{\pi}}{d\tau}=\bar{\pi}\big(1-\alpha_z-3K(0,\tau)\big)-\bar{\pi}^3+\epsilon_a \tag{22}$$

similar to Refs. [34,35].

To derive an equation for $K(\mathbf{s},\tau)$ we consider the time derivative of the average product $\langle\pi(\mathbf{r}_1,\tau)\pi(\mathbf{r}_2,\tau)\rangle$:

$$\frac{\partial}{\partial\tau}\langle\pi(\mathbf{r}_1,\tau)\pi(\mathbf{r}_2,\tau)\rangle=2\bar{\pi}(\tau)\frac{d\bar{\pi}(\tau)}{d\tau}+\frac{\partial K(\mathbf{s},\tau)}{\partial\tau}. \tag{23}$$

On the other hand, it equals

$$\langle \frac{d\pi(\mathbf{r}_1,\tau)}{d\tau}\pi(\mathbf{r}_2,\tau) + \pi(\mathbf{r}_1,\tau)\frac{d\pi(\mathbf{r}_2,\tau)}{d\tau}\rangle. \qquad (24)$$

When substituting Eq. (11) into Eq. (24) and averaging, it is accounted that, for Gaussian random variables, $\langle \xi^3(\mathbf{r}_1,\tau)\xi(\mathbf{r}_2,\tau)\rangle = 3K(0,\tau)K(\mathbf{s},\tau)$ as well as $\langle \xi(\mathbf{r}_1,\tau)\xi^3(\mathbf{r}_2,\tau)\rangle = 3K(0,\tau)K(\mathbf{s},\tau)$, while $\langle \xi(\mathbf{r}_1,\tau)\xi^2(\mathbf{r}_2,\tau)\rangle = 0$. Then, by substituting Eq. (22) into Eq. (23) one obtains finally

$$\frac{dK(\mathbf{s},\tau)}{d\tau} = 2\Delta K(\mathbf{s},\tau) + 2K(\mathbf{s},\tau)[1 - 3\bar{\pi}^2(\tau) - 3K(0,\tau)] + 2\Psi_{zz}(\mathbf{s},\tau). \qquad (25)$$

Equations (21), (22) and (25) present together a closed system of equations which allow determination of functions $\bar{\pi}(\tau)$, $K(\mathbf{s},\tau)$ and $\Psi_{zz}(\mathbf{s},\tau)$. The other correlation functions can be consequently derived from them.

The number of equations and unknown functions can be further reduced by introducing Fourier transforms:

$$K(\mathbf{s},\tau) = \frac{1}{(2\pi)^3}\int d^3q \exp(i\mathbf{q}\mathbf{s})\widetilde{K}(\mathbf{q},\tau), \qquad (26)$$

$$\widetilde{K}(\mathbf{q},\tau) = \int d^3s \exp(-i\mathbf{q}\mathbf{s})K(\mathbf{s},\tau). \qquad (27)$$

In terms of these, Eq. (21) is converted to an algebraic relation

$$q^2\widetilde{\Psi}_{zz}(\mathbf{q},\tau) = -\eta q_z^2 \widetilde{K}(\mathbf{q},\tau). \qquad (28)$$

By applying the Fourier transforms to Eq. (25) and implementing Eq. (28) the cross-correlation function $\widetilde{\Psi}_{zz}$ can be excluded leading to the equation

$$\frac{d\widetilde{K}(\mathbf{q},\tau)}{d\tau} = 2\left[1 - 3\bar{\pi}^2(\tau) - 3K(0,\tau) - \left(q^2 + \eta\frac{q_z^2}{q^2}\right)\right]\widetilde{K}(\mathbf{q},\tau). \qquad (29)$$

Eqs. (22) and (29) form together a closed system of integro-differential equations, since $K(0,\tau)$ is defined by the integral (26). This system of equations will be solved analytically for particular cases and studied numerically for various field and temperature regimes in the next sections.

## 4. Correlation length

Assuming Gaussian properties of the involved random fields [40] allows analytic calculation of the correlation length, a characteristic, well studied experimentally [5-9,13,15,16] and numerically [21] over three decades. To this end we assume an initial isotropic Gaussian form of the correlation function immediately after quenching to the ferroelectric state,

$$K(\mathbf{s},0) = K_0 \exp\left(-\frac{s^2}{2r_c^2}\right) \qquad (30)$$

with an initial polarization dispersion $D(\tau=0) = K(0,0) = K_0$ and the Gauss parameter $r_c$ which will be later related to the initial value of the correlation length. Using Eq. (30) the Fourier transform of the correlation function at $\tau = 0$ can be evaluated from its definition, Eq. (27), as

$$\widetilde{K}(\mathbf{q}, 0) = (2\pi)^{\frac{3}{2}} K_0 r_c^3 \exp\left(-\frac{r_c^2 q^2}{2}\right). \tag{31}$$

To evaluate the time-dependent development of the Fourier transform from its initial value (31) we note that the first order differential equation (29) can be explicitly solved as

$$\widetilde{K}(\mathbf{q}, \tau) = \widetilde{K}(\mathbf{q}, 0) \mu(\tau) \exp\left[-2\left(q^2 + \eta \frac{q_z^2}{q^2}\right)\tau\right] \tag{32}$$

with an auxiliary function of time

$$\mu(\tau) = \exp\{2\tau - 6 \int_0^\tau d\tau' \; [\bar{\pi}^2(\tau') + K(0, \tau')]\} \tag{33}$$

which will be evaluated later.

In terms of the Fourier transform $\widetilde{K}(\mathbf{q}, \tau)$ the correlation length $L(\tau)$ can be defined by a relation [34]

$$L^{-2}(\tau) = \int d^3q \; q^2 \widetilde{K}(\mathbf{q}, \tau) / \int d^3q \widetilde{K}(\mathbf{q}, \tau). \tag{34}$$

Using the presentation (32) the correlation length can represented as

$$L^{-2}(\tau) = \zeta_2(\tau)/\zeta_0(\tau) \tag{35}$$

with auxiliary functions

$$\zeta_2(\tau) = (2\pi)^{-3} \int d^3q \; q^2 \widetilde{K}(\mathbf{q}, 0) \exp\left[-2\left(q^2 + \eta \frac{q_z^2}{q^2}\right)\tau\right] \tag{36}$$

and

$$\zeta_0(\tau) = (2\pi)^{-3} \int d^3q \; \widetilde{K}(\mathbf{q}, 0) \exp\left[-2\left(q^2 + \eta \frac{q_z^2}{q^2}\right)\tau\right]. \tag{37}$$

The integrals in Eqs. (36,37) can be evaluated in a spherical coordinate system in $\mathbf{q}$-space as

$$\zeta_0(\tau) = \frac{1}{2} K_0 r_c^3 \sqrt{\pi/2\eta\tau} \; \text{erf}(\sqrt{2\eta\tau}) \; (r_c^2 + 4\tau)^{-3/2} \tag{38}$$

and

$$\zeta_2(\tau) = \frac{3}{2} K_0 r_c^3 \sqrt{\pi/2\eta\tau} \; \text{erf}(\sqrt{2\eta\tau}) \; (r_c^2 + 4\tau)^{-5/2}. \tag{39}$$

By substituting Eqs. (38,39) in Eq. (35) one obtains

$$L(\tau) = \sqrt{L^2(0) + 4\tau/3} \tag{40}$$

with $r_c^2 = 3L^2(0)$. This analytical form can be compared with experimental results [5-9,13,15,16] as is exemplarily shown in Fig. 2. Experimental data are better fitted by the dependence $L(t) \sim (t-t_0)^\nu$ with $\nu \cong 1/3$ that is discussed in detail below in section 8.

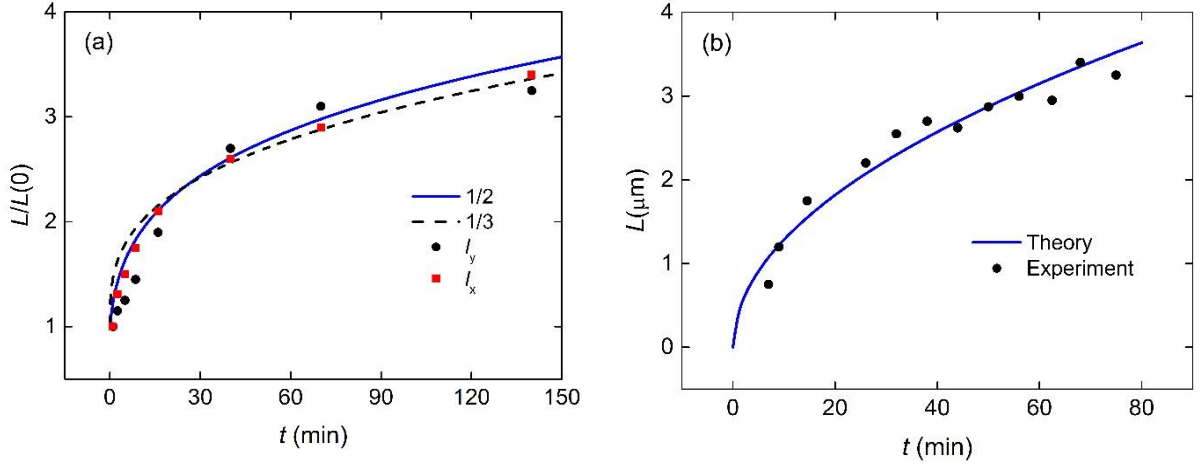

Fig. 2. (a) Correlation length data from Orihara et al. [6] for different in-plane directions perpendicular to polarization in TGS are shown by symbols. The solid blue line presents fitting with Eq. (40) and the black dashed line empirical fitting with the exponent 1/3. (b) Correlation length data for TGS from Golitsyna et al. [16] shown by symbols are fitted with Eq. (40) (solid line). In both cases the trial parameter $L(0) = 1$ was used.

## 5. Phase diagram of asymptotic behavior of the quenched ferroelectric sample

Having established the time-dependent correlation length, a closed system of differential equations for the polarization dispersion (or variance) $D(\tau) = K(0,\tau)$ and the mean polarization $\bar{\pi}(\tau)$ can be derived. To this end we consider Eq. (25) at **s** = 0. First, we note that

$$\Delta K(\mathbf{s}=0,\tau) = -\frac{K(0,\tau)}{L^2(\tau)}. \tag{41}$$

Considering further the term $2\Psi_{zz}(\mathbf{s}=0,\tau)$ we use Eq. (28) to obtain the relation

$$2\Psi_{zz}(\mathbf{s}=0,\tau) = -2\eta K(0,\tau)\zeta_1(\tau)/\zeta_0(\tau) \tag{42}$$

with an auxiliary function

$$\zeta_1(\tau) = (2\pi)^{-3}\int d^3q \frac{q_z^2}{q^2} \widetilde{K}(\mathbf{q},0) \exp\left[-2\left(q^2+\eta\frac{q_z^2}{q^2}\right)\tau\right] \tag{43}$$

which can be evaluated as

$$\zeta_1(\tau) = \frac{1}{2}K_0 r_c^3 \frac{1}{2\eta\tau}\left[\frac{\sqrt{\pi}}{2\sqrt{2\eta\tau}}\operatorname{erf}\left(\sqrt{2\eta\tau}\right) - \exp(-2\eta\tau)\right](r_c^2+4\tau)^{-3/2}. \tag{44}$$

Then Eq. (25) at **s** = 0 can be transformed to

$$\frac{dD(\tau)}{d\tau} = [2 - 6\bar{\pi}^2(\tau) - 6D(\tau) - \nu(\tau)]D(\tau) \tag{45}$$

with an auxiliary function

$$v(\tau) = \frac{2}{L^2(\tau)} + \frac{1}{2\tau}\left[1 - \frac{2}{\sqrt{\pi}} \frac{\sqrt{2\eta\tau}\exp(-2\eta\tau)}{\text{erf}(\sqrt{2\eta\tau})}\right]. \tag{46}$$

A comparison with previous studies reveals that the last term in Eq. (46) uniquely represents the effect of depolarization fields neglected before [34-37].

By dividing Eq. (45) with $D(\tau)$ and integrating, a useful relation for the auxiliary function $\mu(\tau)$, Eq. (33), can be established:

$$\mu(\tau) = \frac{D(\tau)}{D(0)}\exp\left(\int_0^\tau d\tau'\, v(\tau')\right). \tag{47}$$

This can be further evaluated by substituting $v(\tau)$ from Eq. (46) in Eq. (47) to obtain

$$\mu(\tau) = \frac{D(\tau)}{D(0)}\left(1 + \frac{4\tau}{r_c^2}\right)^{3/2} \frac{2}{\sqrt{\pi}} \frac{\sqrt{2\eta\tau}}{\text{erf}(\sqrt{2\eta\tau})}. \tag{48}$$

Thus, the problem of finding the correlation function $\widetilde{K}(\mathbf{q},\tau)$, Eq. (32), is reduced to the finding of the dispersion $D(\tau)$.

The latter can be determined by solving Eq. (45) together with equation (22), rewritten as

$$\frac{d\bar{\pi}(\tau)}{d\tau} = \bar{\pi}(\tau)\bigl(1 - \alpha_z - 3D(\tau)\bigr) - \bar{\pi}^3(\tau) + \epsilon_a. \tag{49}$$

A closed system of differential equations (45) and (49) for $D(\tau)$ and $\bar{\pi}(\tau)$ will be studied in the following numerically. First, however, we will investigate equilibrium points of these equations under the asymptotic condition $\tau \to \infty$ that allows the formulation of a phase diagram in terms of $\bar{\pi}$ and $D$.

The solution of the system of evolution equations (45, 49) for average polarization $\bar{\pi}$ and its dispersion $D$ with given initial conditions ($\bar{\pi}(0) = \pi_0$, $D(0) = D_0$) provides information about the development of domain structures and the final stages of the ordering process. We note that the parameter $\alpha_z = 2\chi_f \rho_z = 2\rho_z C/(T_c - T)$, where $C$ is the Curie constant of a material, is both temperature and geometry dependent via Eq. (8) and can vary in a wide range. On the one hand, it can be strongly decreased to $\alpha_z \ll 1$ by reducing the dielectric layer thickness $h_d$; on the other hand, for any fixed geometry, it can be strongly enhanced to $\alpha_z \gg 1$ by approaching the transition temperature $T_c$, limited, however, by the criterion of the applicability of the LGD theory (see Appendix A). The other variable parameter in equations (45,49), $\epsilon_a$, is the externally induced mean field in the ferroelectric normalized to the very large field $E_0$, which is comparable with the thermodynamic coercive field. Thus, the interplay between the single-domain and multi-domain states is expected at fields $\epsilon_a < 1$.

Consideration of ordering processes at large times suggests that the right-hand side of equations of the system (45,49) vanishes according to $\partial \bar{\pi}/\partial \tau \to 0$, $\partial D/\partial \tau \to 0$. This reduces the system of evolution equations to an algebraic system which can be qualitatively analyzed using the phase diagram concept [41]:

$$\begin{cases}(1 - \alpha_z - 3D)\bar{\pi} - \bar{\pi}^3 + \epsilon_a = 0 \\ (1 - 3\bar{\pi}^2 - 3D)D = 0.\end{cases} \tag{50}$$

There are six equilibrium points in the ferroelectric phase ($\alpha_z > 0, T < T_c$) in the phase diagram on the $(\bar{\pi}, D)$-plane (Fig. 3) with coordinates:

$$\begin{cases}\bar{\pi}_1 = \frac{2}{\sqrt{3}}\sqrt{1 - \alpha_z} \cdot \cos\left(\frac{1}{3}\arccos\frac{3\sqrt{3}\,\epsilon_a}{2(1-\alpha_z)^{\frac{3}{2}}} + \frac{2\pi n}{3}\right) \\ \text{(I):} \quad \bar{\pi} = \bar{\pi}_1, \; n = n_1 = 2 + 3(k-1), k \in Z; \quad D = 0 \\ \text{(II):} \quad \bar{\pi} = \bar{\pi}_1, \; n = n_2 = 3(k-1), k \in Z; \quad D = 0 \\ \text{(III):} \quad \bar{\pi} = \bar{\pi}_1, \; n = n_3 = 1 + 3(k-1), k \in Z; \quad D = 0\end{cases} \tag{51a}$$

$$\begin{cases}\bar{\pi}_2 = \frac{2}{\sqrt{6}}\sqrt{\alpha_z} \cdot \cos\left(\frac{1}{3}\arccos\left(-\frac{3\sqrt{6}\,\epsilon_a}{2\alpha_z^{3/2}}\right) + \frac{2\pi n}{3}\right) \\ \text{(IV):} \quad \bar{\pi} = \bar{\pi}_2 \text{ with } n = n_1; \quad D = 1/3 - \bar{\pi}_2^2 \\ \text{(V):} \quad \bar{\pi} = \bar{\pi}_2 \text{ with } n = n_2; \quad D = 1/3 - \bar{\pi}_2^2 \\ \text{(VI):} \quad \bar{\pi} = \bar{\pi}_2 \text{ with } n = n_3; \quad D = 1/3 - \bar{\pi}_2^2\end{cases} \tag{51b}$$

The equilibrium point I is an unstable node corresponding to the unpoled state. The system falls into the vicinity of this point after the quenching from the paraelectric phase into the ferroelectric one. This is a starting point of domain growth and evolution to one of thermodynamically (meta-)stable states – the nodes II, III, and IV. Points II and III with $D = 0$ correspond to the formation of single-domain states with polarization vectors directed along and opposite to the applied field, respectively. Point IV has a remarkable dispersion value and corresponds to the formation of a thermodynamically (meta-)stable multi-domain state of quasi-periodic type [13] with 180° domain walls.

The evolution of ferroelectric domain structures in time can proceed non-monotonically with a formation of short-lived intermediate states corresponding to the saddle points V and VI. These points are distinguished by a pronounced asymmetry of the volume fractions of domains of the opposite direction of polarization (depending on the applied field). Phase trajectory of the domain structure evolution can pass very close to the saddle point, but never reach it, and continue its evolution after a short-lived slowing down to the (meta-)stable states (II, III, or IV). Points V and VI are intersection nodes of the separatrices dividing the phase diagram into three sectors characterizing the "attraction areas" of (meta-)stable single-domain (*1* and *2*) and multi-domain (*3*) states (Fig. 3).

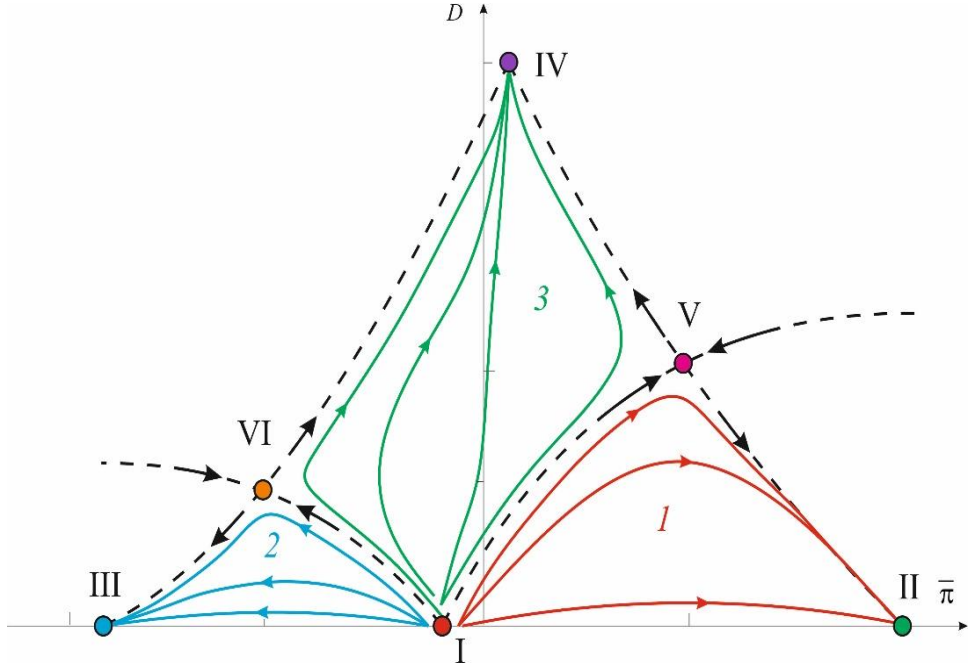

Fig. 3. Schematic view of the phase diagram of the system in variables $(\bar{\pi}, D)$ at exemplary values of $\alpha_z = 1/2$ and $\epsilon_a = 0.002$ with indication of equilibrium points (Roman numerals), separatrices (dashed lines), and sectors of attraction of thermodynamically (meta-)stable states: single-domain (*1* and *2*) and multi-domain (*3*). Arrows show the direction of entry and exit of separatrices in saddle points V and VI. Solid lines show the variety of possible phase trajectories and ways of the domain structure evolution to the states of thermodynamic equilibrium or metastability.

It is seen from Eqs. (51) that the positions of equilibrium points in the phase diagram significantly depend on the values of $\alpha_z$ (controlled by the geometry and temperature) and the field $\epsilon_a$. In the absence of the external electric field the phase diagram is symmetrical (Fig. 4). Point I is located exactly in the origin, point IV does not have a displacement on the polarization axis, and points II and III as well as V and VI are located symmetrically to each other. However, the presence of even a very weak external electric field remarkably shifts the position of equilibrium points in the phase diagram. With an increase in electric field, equilibrium points continue to move and can disappear at some critical field values. These values can be obtained from the analysis of coordinates of equilibrium points (51) or directly from the system (50).

For a single-domain state with $D = 0$, the electric field can be expressed from the first equation (50) as $\epsilon_a = \bar{\pi}^3 - (1 - \alpha_z)\bar{\pi}$, where two extremes in symmetric points exist, $\bar{\pi}_c^{(1,2)} = \pm\left(\frac{1-\alpha_z}{3}\right)^{1/2}$. As a result, a minimal $\epsilon_s^{min}$ and a maximal $\epsilon_s^{max}$ critical electric fields can be obtained for positive and negative order parameters $\bar{\pi}$, limiting their existence to a field region around $\epsilon_a = 0$,

$$\epsilon_s^{min} = -\frac{2}{3\sqrt{3}}(1-\alpha_z)^{\frac{3}{2}} < \epsilon_a < \frac{2}{3\sqrt{3}}(1-\alpha_z)^{\frac{3}{2}} = \epsilon_s^{max}. \tag{52}$$

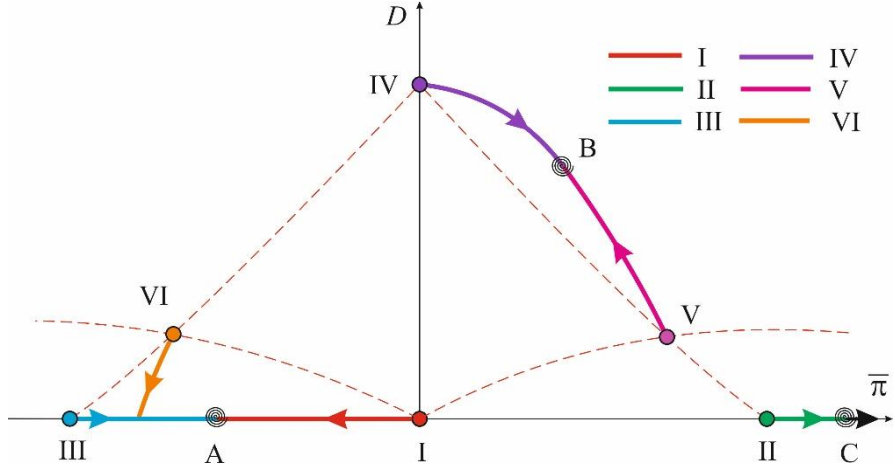

Fig. 4. Transformation of the phase diagram of the system in variables $(\bar{\pi}, D)$ (Fig. 3) upon the smooth increase of the external electric field $\epsilon_a$ at an exemplary value $\alpha_z = 1/2$ with an indication of equilibrium points (Roman numerals). Red dashed lines indicate the separatrices dividing the symmetrical phase diagram in the absence of electric field, $\epsilon_a = 0$. Arrows show the displacement of equilibrium points with increasing electric field. Point A indicates the place of convergence of the equilibrium points I and III which occurs at the field value $\epsilon_a = \epsilon_s^{max}$, whereas point B corresponds to the convergence of the equilibrium points IV and V which occurs at $\epsilon_a = \epsilon_m^{max}$.

The existence of stable and metastable states II and III is possible in the said region only if the parameter $\alpha_z < 1$, as is apparent from Eqs. (51a) and (52). Particularly, in the case $\alpha_z = 0$ critical fields coincide with the thermodynamical coercive fields [39]. In contrast, for $\alpha_z > 1$, only the point II persists as a possible single-domain state for all positive fields, while the point III persists as the only possible single-domain state for all negative fields.

For a multi-domain state with $D \neq 0$, the dispersion can be expressed from the second equation (50) as $D = 1/3 - \bar{\pi}^2$ resulting in the expression for the electric field from the first equation (50), $\epsilon_a = \alpha_z \bar{\pi} - 2\bar{\pi}^3$. The analysis of the latter cubic equation reveals an existence of the multi-domain state IV, as well as of the two saddle points V and VI, in the field region

$$\epsilon_m^{min} = -\frac{2}{3\sqrt{6}}\alpha_z^{3/2} < \epsilon_a < \frac{2}{3\sqrt{6}}\alpha_z^{3/2} = \epsilon_m^{max}, \tag{53}$$

which is valid for any positive $\alpha_z$. Beyond this field region only single-domain states may exist.

Considering the smooth increase of the electric field $\epsilon_a$ the motion of equilibrium points in the phase diagram can be observed (arrows in Fig. 4). When the value of the electric field becomes critical, $\epsilon_a = \epsilon_s^{max}$, points I and III converge at the point A, similarly to previous studies [42]. In the same way, points IV and V converge at a point B at the field value $\epsilon_a = \epsilon_m^{max}$. The succession of these events depends on the value of $\alpha_z$. When an electric field exceeds both values $\epsilon_s^{max}$ and $\epsilon_m^{max}$, the sectors 2 and 3 in the phase diagram disappear (Fig. 3). Then only the point II continues to move to the right with the increasing field (Fig. 4). Thus, a high value of the electric field inhibits the intermediate stages (points V and VI), and the phase diagram consists of one sector *1* (Fig. 3), so that the only one single-domain state in point C directed along the electric field exists in Fig. 4.

A phase diagram in terms of the electric field and the parameter $\alpha_z$, depending in turn on the material parameters, structure geometry, and temperature, summarizes all possible states in Fig. 5. The region of the possible existence of (meta-)stable single-domain states II and III as well as of the unstable point I is delineated by red solid lines. In the region delineated by green solid lines the single-domain state II or the multi-domain state IV can be realized for positive electric fields, while the single-domain state III or the multi-domain state IV can be realized for negative electric fields. In the area of overlapping of the two mentioned regions, the existence of both (meta-)stable single-domain states II and III, and the multi-domain state IV is possible, while the unstable equilibrium point I and the saddle points V and VI are available too. Outside the said regions only the single-domain state II can exist for positive and the single-domain state III for negative electric fields. We note, however, that introducing even a slight inhomogeneity of the ferroelectric is known to lead to the stabilization of the multi-domain state [43].

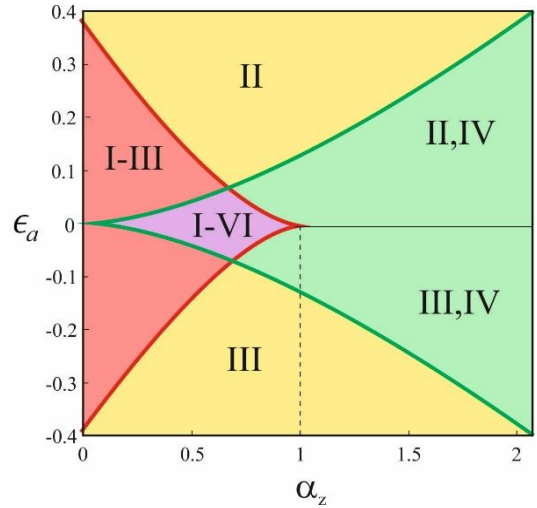

Fig. 5. Phase diagram of possible states on the parameter plane $(\epsilon, \alpha_z)$.

It is known that the physical reason of the multi-domain state formation is the reduction of the energy accumulated in depolarization fields [44,45]. The stabilizing effect from the spatial variation of the polarization can be observed in the fluctuation contribution to the energy (1) which amounts to $\Delta \Phi \sim - |A|P_s^2 D^2(\tau \to \infty) V_f$ as is shown in more detail in Appendix C.

## 6. Temporal behavior of the ordering process

The nonlinearity of the obtained evolution equations (45,49) does not allow to describe all stages of the ordering process analytically. Therefore, to trace in detail the evolution of the domain structure towards the thermodynamic equilibrium, a numerical analysis of the system of differential equations (45,49) with certain initial conditions $(\bar{\pi}_0, D_0)$ was carried out in MATLAB package. The ferroelectric sample in this study is considered to be quenched without any external electric field. Thus, the average polarization of the crystal is zero in the initial stage $\bar{\pi}_0 = 0$ after quenching and changes in time during relaxation when the external electric field is turned on. However, small polarization inhomogeneities may emerge in the nonequilibrium system during the quench and become a certain size by the time relaxation begins, characterized by some initial values of dispersion $D_0 \neq 0$ and the correlation length $L(0)$. The process of domain nucleation in a quenched sample is a random one and can hardly be controlled experimentally. But the values of quenching temperature and external electric field still can be used to manage the domain growth and direct it to the formation of a certain type of domain structure. The other parameters used in the calculation were $\eta = 1$ and $\alpha_z = 1/2$ chosen from the region I-VI in the phase diagram of Fig. 5 which exhibits the richest variety of evolution scenarios.

## A. Influence of external electric field on the domain structure formation

The influence of an external electric field on the ordering process in the ferroelectric was analyzed for a constant finite isothermal exposure of the sample. It was shown that varying the magnitude of the electric field can change not only the phase trajectory of the evolution of the system but also its final result. In the absence of the external electric field, the domains of both polarization directions gradually arise and monotonously grow (curve 1 in Fig. 6) with the formation of a stable multi-domain structure corresponding to the point IV in the phase diagram of Fig. 3. In the case of a TGS crystal which has 180° domain walls, this state is a one-dimensional quasi-periodic domain structure of irregular stripes which clearly exhibits a principal mode by the Fourier analysis of experimental data [13]. When a positive electric field is imposed on the sample some deviation of the average polarization in the direction of this field is observed (curve 2 in Fig. 6) so that, in the end, a stable multi-domain structure is formed in the system. The curves 3 and 4 in Fig. 6 are of particular interest because they run in the area of separation of single-domain (*1* in Fig. 3) and multi-domain (*3* in Fig. 3) ordering regions. Compared to the curves 1, 2, 5, and 6 in Fig. 6 which pass far from the saddle point V (Fig. 3), the phase trajectories 3 and 4 pass along the separatrix on opposite sides of it and fall into the vicinity of the saddle point. This is a region where the electric field can drastically change both the further evolution of the system and the type of the final domain structure. This can be interpreted as hitting a bifurcation point where the slightest variation of the magnitude of the electric field can change the trend of the further ordering of the system from a multi-domain state (curve 3 in Fig. 6) to a single-domain one (curve 4 in Fig. 6).

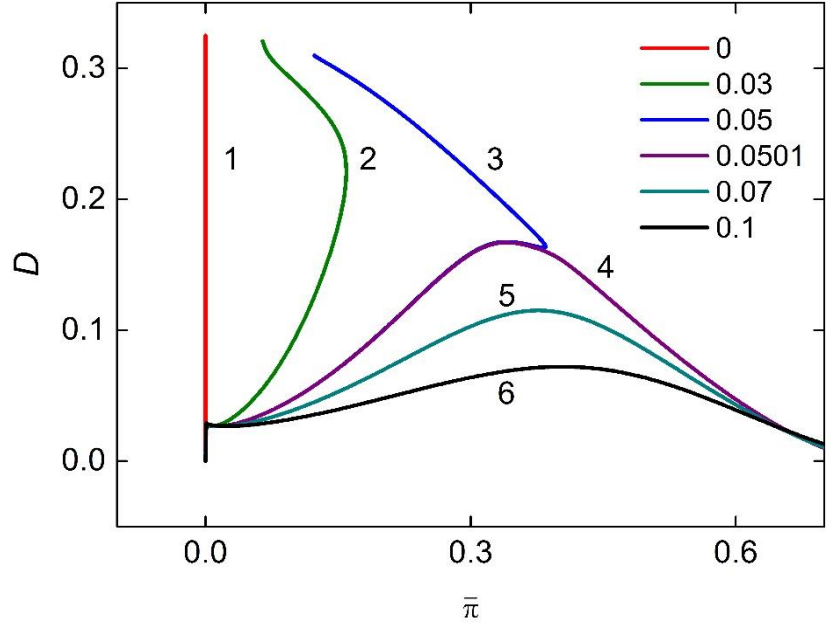

Fig. 6. Phase trajectories of the system evolution calculated for the parameters: $\bar{\pi}_0 = 0$, $D_0 = 0.0001$, $L(0) = 1$, $\alpha_z = 1/2$, $\eta = 1$. Curves 1–6 correspond to the values of the external electric field in the sample $\epsilon_a = \{0; 0.03; 0.05; 0.0501; 0.07; 0.1\}$, respectively.

The vicinity of the saddle point V (Fig. 3) can be described as a region of the kinetic slowing down where the relaxation process is retarded for a while (Fig. 7). The dynamics of domain growth and rearrangement hang on here, and the further scenario of the system evolution has a probabilistic character. It is important to note that the domain structure, which had already formed by the time the system enters the intermediate phase, is characterized by a pronounced asymmetry in the volume fractions of domains of different signs. This means that the width of domains directed along the external electric field is much larger than the width of the reversely directed domains. However, the further evolution of the system to the state of thermodynamic equilibrium still may proceed either as the further growth of the prevailing domains (curve 4 in Fig. 7) or their decrease with a tendency to establish a balance between the domains of opposite signs (curve 3 in Fig. 7).

The duration of kinetic slowing down of the system near the saddle point V (Fig. 3) can be estimated from the length of the plateau or step (curves 3,4) on the time evolution curves for the average polarization $\bar{\pi}$ (Fig. 7($a$)) and its dispersion $D$ (Fig. 7($b$)). The appearance of an intermediate stage during the evolution of the system significantly slows down the overall process of ordering and delays the onset of thermodynamic equilibrium. Evolution plots (Fig. 7) show that phase trajectories passing far from the saddle point come to the stable multi-domain (curves 1, 2) and single-domain (curves 5, 6) states much faster than curves 3 and 4, respectively.

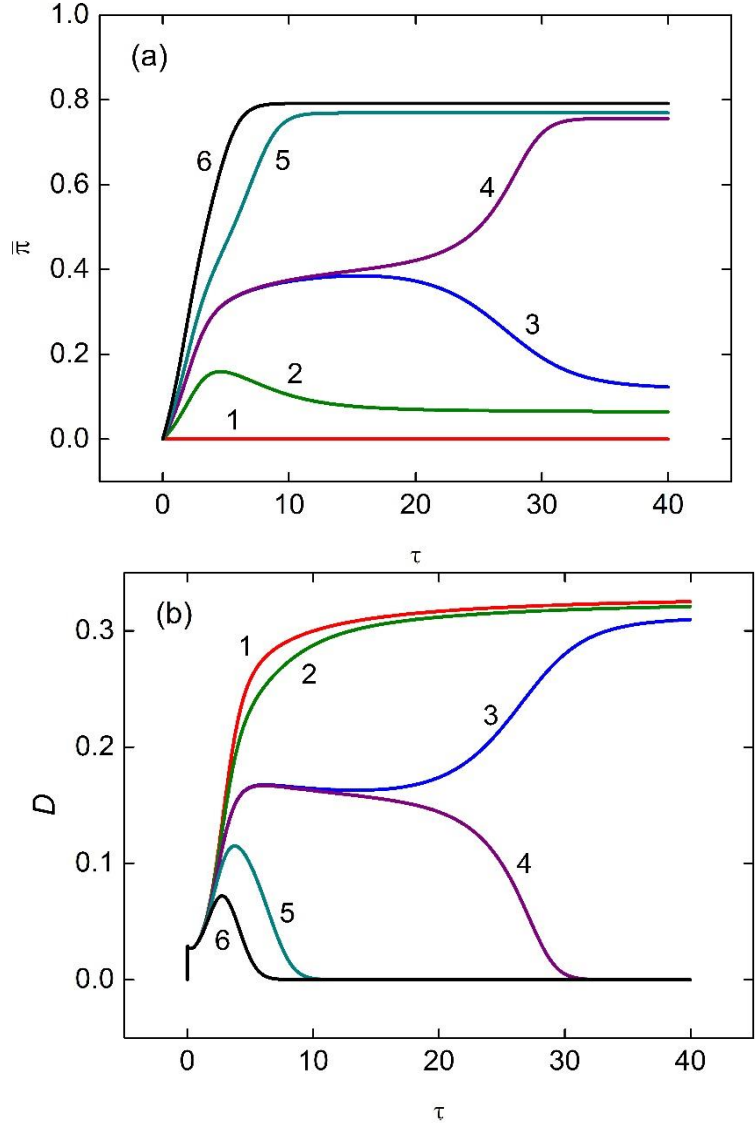

Fig. 7. (a) Evolution curves for average polarization $\bar{\pi}$ and (b) its dispersion $D$ calculated for the same parameter values as in Fig. 6 against dimensionless time $\tau$.

### B. Influence of polarization-induced electric fields on the domain structure formation

The effect of depolarization fields, induced by stochastic nonuniform polarization development, on the ordering kinetics can be studied by formal setting the parameter $\eta = 0$ in Eq. (45). Then the second term in the auxiliary function $v(\tau)$, Eq. (46), describing depolarization fields, disappears and the results can be compared with those in Figs. 6 and 7 with $\eta = 1$. It turned out that depolarization fields crucially affect the time development of domain structures though the final states in Figs. 3 and 4 are independent of $\eta$. A comparison of phase trajectories calculated at the same parameters and values of the applied field with an account of depolarization fields and without it demonstrates the tendency to the multi-domain states at low applied fields in the first case (curves 2 and 3 in Figs. 6-7) and the tendency to the single-domain state for the same field values in the second case (curves 1 and 2 in Fig. 8(a)). In the absence of the stochastic depolarization fields, the system develops towards the multi-domain state first at a much lower field of $\epsilon_a = 0.015$. (curves 5 in Fig. 8).

Thus, the depolarization fields significantly contribute to the system response to an external electric field. Trying to compensate for it, they prevent a rapid orientation of domains along the external electric field, retaining the tendency to form a multi-domain structure. Considering the phase diagram in Fig. 5, the switching to the multi-domain state occurs well below the boundaries of the region I-VI if the stochastic fields are neglected. With an account of them, the switching to the multi-domain state occurs virtually immediately when crossing the boundaries towards the region I-VI.

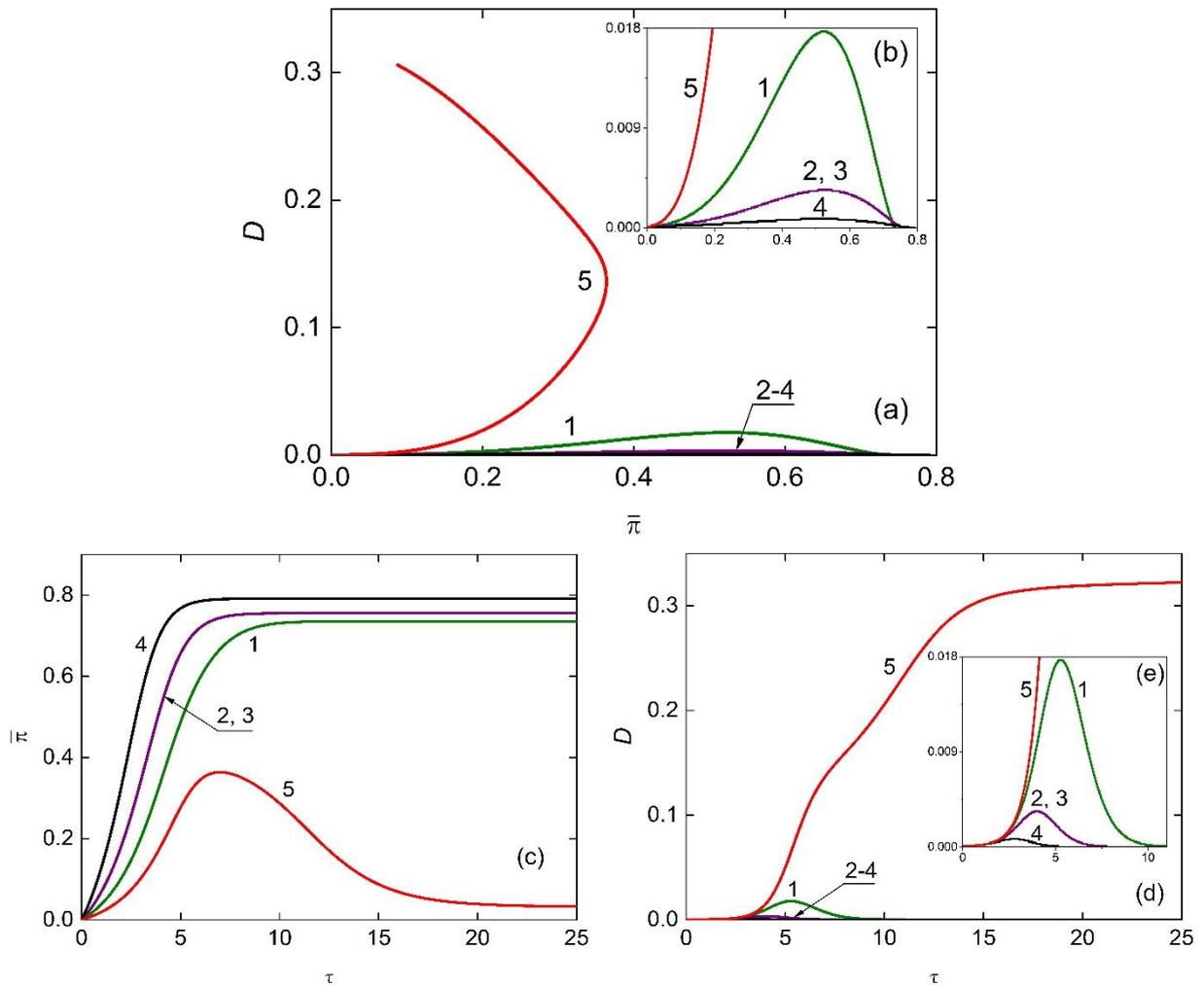

Fig. 8. (a) Phase trajectories of the system without the effect of depolarization fields ($\eta = 0$). Curves 1–4 correspond to the values of the external electric field in the sample $\epsilon_a = \{0.03; 0.05; 0.0501; 0.07\}$, respectively (cf. Fig. 7), while the curve 5 is for the field $\epsilon_a = 0.015$. The insert (b) shows the upscaled curves 1-4. (c) Evolution curves for average polarization $\bar{\pi}$ and (d) polarization dispersion $D$ in the absence of depolarization fields were calculated for the same parameters vs. dimensionless time $\tau$. The inset (e) shows the upscaled curves 1–4.

## 7. Longitudinal and transverse polarization correlations

Having established numerically the behavior of the dispersion $D(\tau)$ the correlation function $K(\mathbf{s}, \tau)$ can be derived from Eqs. (26), (32) and (48) as

$$K(\mathbf{s}, \tau) = \mu(\tau) \frac{1}{(2\pi)^3} \int d^3q \, \widetilde{K}(\mathbf{q}, 0) \exp\left(i\mathbf{q}\mathbf{s} - 2(q^2 + \eta \frac{q_z^2}{q^2})\tau\right). \tag{54}$$

Now, by substituting $\widetilde{K}(\mathbf{q}, 0)$ from Eq. (31) in Eq. (54) one finds a general expression

$$K(\mathbf{s}, \tau) = \frac{\mu(\tau) K_0 r_{c0}^3}{(2\pi)^{3/2}} \int d^3q \exp\left(i\mathbf{q}\mathbf{s} - q^2\left(\frac{r_c^2}{2} + 2\tau\right) - \frac{q_z^2}{q^2} 2\eta\tau\right). \tag{55}$$

We note that, in experiments [5-9,16], not the correlation function but the normalized correlation coefficient is actually measured, which can be expressed as $C(\mathbf{s}, \tau) = K(\mathbf{s}, \tau)/D(\tau)$. Below this quantity is evaluated separately for the cases of longitudinal correlations along the polarization direction, $\mathbf{s} = (0, 0, s_z)$, and transverse correlations in the plane perpendicular to the polarization, $\mathbf{s} = (\mathbf{s}_\perp, 0)$, with $\mathbf{s}_\perp = (s_x, s_y)$, typically studied experimentally.

We start with the longitudinal correlations reducing the correlation coefficient to

$$C_\parallel(s_z, \tau) = \frac{\mu(\tau) K_0 r_c^3}{D(\tau)(2\pi)^{1/2}} \int_0^\infty dq q^2 \exp\left(-q^2\left(\frac{r_c^2}{2} + 2\tau\right)\right)$$

$$\times \int_0^\pi d\theta \sin\theta \exp(iqs_z \cos\theta - 2\eta\tau \cos^2\theta). \tag{56}$$

By integration and using the explicit formula for $\mu(\tau)$, Eq. (48), this function can be obtained in a closed form

$$C_\parallel(s_z, \tau) = \frac{\sqrt{2\eta\tau}}{\text{erf}(\sqrt{2\eta\tau})} \left\{ \frac{\text{erf}\sqrt{2\eta\tau + s_z^2/(6L^2(\tau))}}{[2\eta\tau + s_z^2/(6L^2(\tau))]^{3/2}} 2\eta\tau + \frac{2}{\sqrt{\pi}} \frac{s_z^2}{6L^2(\tau)} \frac{\exp[-2\eta\tau - s_z^2/(6L^2(\tau))]}{[2\eta\tau + s_z^2/(6L^2(\tau))]} \right\} \tag{57}$$

with the correlation length $L(\tau)$ defined by Eq. (40).

We continue now with the transverse correlations by reducing Eq. (55) to

$$C_\perp(\mathbf{s}_\perp, \tau) = \frac{2}{\sqrt{\pi}} \frac{\sqrt{2\eta\tau} \exp(-2\eta\tau)}{\text{erf}(\sqrt{2\eta\tau})} \int_0^1 dz \frac{1}{\sqrt{1-z}} \exp[-z(u - 2\eta\tau)]$$

$$\times \left[\left(1 - \frac{u}{2}z\right) I_0(uz) + \frac{u}{2} z I_1(uz)\right] \tag{58}$$

with the modified Bessel functions $I_0(uz)$ and $I_1(uz)$, where a combined variable $u = s_\perp^2/(12L^2(\tau))$ was introduced for convenience. Interestingly, in the limit of the absence of field-mediated correlations, which is formally realized at $\eta \to 0$, the correlation coefficient can be calculated in the closed form for arbitrary $\mathbf{s}$ and equals simply

$$C(\mathbf{s}, \tau) = \exp\left(\frac{-s^2}{6L^2(\tau)}\right) \tag{59}$$

retaining thus the initial Gaussian form, Eq. (30), with the time-dependent width.

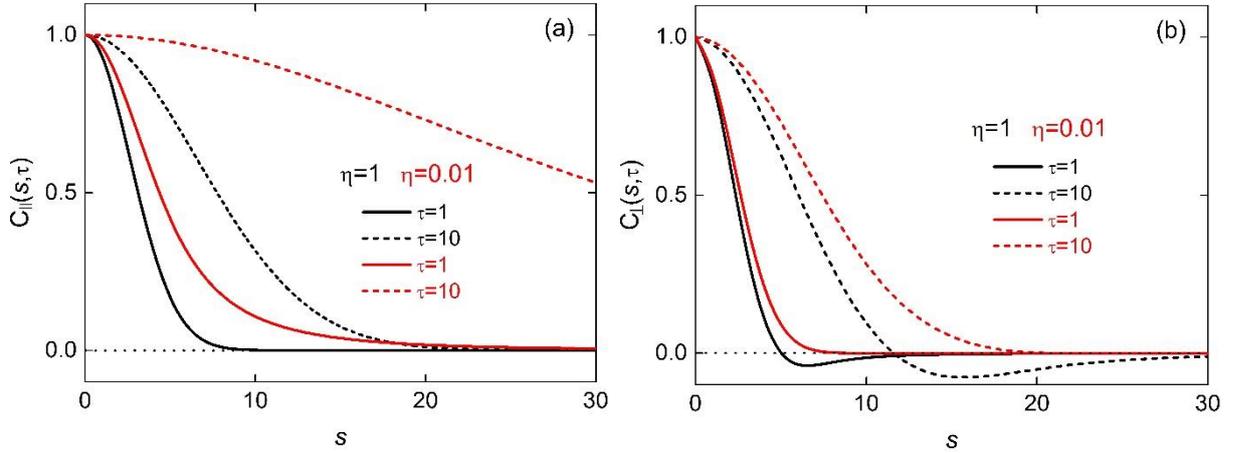

Fig. 9. Spatial dependence of the longitudinal (a) and the transverse (b) correlation coefficients of polarization for a substantial ($\eta = 1$, black lines) or infinitesimal ($\eta = 0.01$, red lines) normalized ferroelectric susceptibility, corresponding to the neglection or the account of the depolarization field effect, evaluated exemplarily for two dimensionless times, $\tau = 1$ and $\tau = 10$.

Spatial dependences of correlation coefficients according to the analytical formulas (57) and

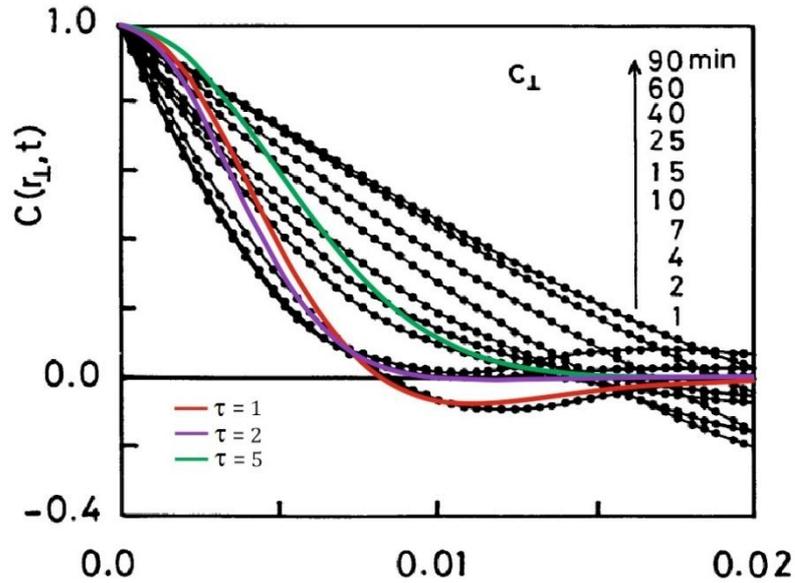

Fig. 10. Spatial dependences of the "transverse" correlation coefficient of polarization in TGS for different times from the work by Tomita et al. [5](© (1989) The Physical Society of Japan) are shown by black dots and lines. Colored lines exhibit the transverse correlation coefficient $C_\perp(\mathbf{s}_\perp, \tau)$, Eq. (58) evaluated for different dimensionless times $\tau = 1, 2$ and $5$ as indicated on the plot.

(58) are presented for different times $\tau$ in Fig. 9. Note the Gaussian spatial dependence of both correlation coefficients for $\eta \ll 1$. Furthermore, for the longitudinal correlation coefficient, Eq. (57), an approximate Gaussian behavior is observed for all times at arbitrary $\eta$ in Fig. 9(a). In contrast, the transverse coefficient, Eq. (58), is changing its sign, as is seen in Fig. 9(b), and thus exhibits clearly non-Gaussian behavior.

A comparison of the spatial dependence of the transverse correlation coefficient $C_\perp(\mathbf{s}_\perp, \tau)$ with experimental data by Tomita et al. [5] is shown in Fig. 10 for different times. Note the distinct definitions of the "transverse" and "longitudinal" correlation coefficients in experimental works [5-9] referred to the orientations with respect to the domain boundaries at the surface of TGS crystals.

## 8. Discussion and conclusions

In this work we advanced a self-consistent theory of stochastic development of polarization domain structures in a uniaxial ferroelectric/nonferroelastic starting from an unpoled state obtained by quenching from a high-temperature paraelectric phase. The model is based on the LGD approach and accounts for the effect of an applied electric field on the system evolution as well as for the feedback via depolarization electric fields induced by the emerging domain structure. The polarization and electric field components are treated as Gaussian random variables since the structure emerges from an initial unpoled disordered state. The effect of thermodynamic fluctuations is assumed to be negligible which is justified as soon as the spatial scale of initial disorder exceeds that of thermodynamic fluctuations and the Levanyuk criterion of the applicability of the LGD theory is satisfied, as is shown in Appendix A. In the case of TGS this means that the model applies at temperatures $|T - T_c| > 0.02\, K$.

A closed system of integro-differential equations (22,29) was formulated for the time-dependent two-site correlation function $K(\mathbf{s}, \tau)$ for polarization and the time-dependent mean polarization $\bar{\pi}(\tau)$. The other correlation functions can be derived from the said functions. Considering the polarization dispersion $D(\tau) = K(\mathbf{s}=0, \tau)$ reduces the problem to a closed system of differential equations (45,49) for $D(\tau)$ and $\bar{\pi}(\tau)$. The study of the field-dependent equilibrium points of the latter system allows a construction of a phase diagram (Fig. 3) of possible stable and metastable states of the system in coordinates $(\bar{\pi}, D)$. Considering the effect of the applied field $\epsilon_a$ on the above phase diagram (Fig. 4) reveals areas of possible stable and metastable states on the plane $(\alpha_z, \epsilon_a)$ (Fig. 5) where the combined characteristic $\alpha_z$ depends on the system geometry, material parameters, and temperature. Particularly, the field region of the existence of multi-domain states is reduced, when $\alpha_z \ll 1$ together with the dielectric layer thickness $h_d \ll h_f$, and enhanced, when $\alpha_z \gg 1$ due to approaching the phase transition temperature $T_c$.

A numerical solution of the equations (45,49) exhibits the time development of the dispersion $D(\tau)$ and the mean polarization $\bar{\pi}(\tau)$ and reveals a sharp bifurcation behavior from multi-domain towards single-domain states when increasing the applied field (Fig. 6). This is accompanied by non-monotonic time dependences of both $D(\tau)$ and $\bar{\pi}(\tau)$ as well as the slowing down development in the bifurcation regime (Fig. 7) corresponding to the passing by saddle points in the phase diagram of Fig. 3. Comparing the time trajectories for $D(\tau)$ and $\bar{\pi}(\tau)$ in Fig. 8 in the presence and in the absence of the stochastic depolarization fields reveals the crucial role of the latter for the domain kinetics. Depolarization fields cannot change the equilibrium points in the phase diagram (and thus the available final states of the evolution) but affect the domain formation kinetics and the choice of a final state at a certain applied field value. Thus, the account of the stochastic depolarization fields supports the existence of multi-domain states in a substantially wider applied field region.

Assuming an initial Gaussian shape of the correlation function $K(\mathbf{s}, \tau = 0)$, Eq. (30), allows an exact calculation of the correlation length $L(\tau)$ defined by Eq. (34). It exhibits a simple diffusion-like behavior, Eq. (40). Though it roughly captures the time development of the correlation length $L(t) \sim (t - t_0)^\nu$ observed in experiments, as is shown in Fig. 2, the experimental data are better described by the exponent $\nu$ about 0.2-0.3 [5-10,16]. Furthermore, an observed linear time behavior for temperatures close to $T_c$ [3,16] is missing in the theory as well as the asymptotic dependence $L(t) \sim [\ln(t/t_0)]^4$ observed at longer times [7-9]. The latter behavior was explained theoretically [21] as an effect of the frozen random fields due to the presence of defects which are not accounted in our model.

Though the correlation function $K(\mathbf{s}, \tau)$ can only be evaluated numerically from the integro-differential equations (22,29), the correlation coefficient $C(\mathbf{s}, \tau) = K(\mathbf{s}, \tau)/D(\tau)$, which is actually measured in experiments [5-9,16], can be found analytically. For the longitudinal correlations along the polarization direction a closed formula (57) was obtained, while for the correlations in the plane perpendicular to the polarization direction (coinciding with the surface of the crystal) an integral expression (58) was derived. The weakness of the current theory consists in the assumption of isotropic correlations in the said plane, while in the available experiment on TGS [5-9,16] the correlations are strongly anisotropic like the crystal itself. For that reason, the theory is still unable to explain the observed oscillations in the correlation coefficient perpendicular to the domain boundaries in the plane surface [5-9]. However, the correlations along the domain boundaries in the plane are roughly captured by Eq. (58) as is shown in Fig. 10. Nevertheless, the initial linear spatial dependence $C(\mathbf{s}, \tau) \cong 1 - s/L(\tau)$ typically observed in experiment [7-9,16] is not explained by the theory which assumes the Gaussian dependence. Also, the hypothesis of scaling dependence of $C(\mathbf{s}, \tau)$, supported at $\eta = 0$ by Eq. (59), is not generally confirmed by the theory for arbitrary $\eta$, since the correlation coefficients in both directions, Eq. (57) and (58), contains both the scaling forms of the type $s/L(\tau)$ and an explicit time dependence. Furthermore, we note the decoupling of the monotonic time dependence of the correlation length

$L(t)$ and a generally non-monotonic time dependence of the dispersion $D(\tau)$ which, however, does not contain a spatial scale.

In conclusion, for a better description of the available experiments an anisotropic LGD thermodynamic potential (1) is required, as well as a non-Gaussian initial shape of the correlation function and probably a consideration of non-Gaussian random variables.

## Appendix A. Limitations on the application of the model

In the model presented in Section 2, prevailing of quenched disorder over thermodynamic fluctuations is assumed. Here we identify the parameter range where this assumption is valid. Spatial variations of polarization due to the quenched disorder and due to the thermodynamic fluctuations are both short-range ones with different characteristic lengths. The quenched disorder is characterized initially by the correlation function (30) and can be represented in physical units as

$$\langle \Delta P(\mathbf{r}+\mathbf{s})\Delta P(\mathbf{s})\rangle = D(0)P_s^2 \exp\left(-\frac{s^2}{6L^2(0)}\right), \qquad (A1)$$

where the spatial scale of polarization fluctuations is, in principle, time-dependent, Eq. (40), and temperature independent. The thermodynamic fluctuations can be characterized by the correlation function, derived within the Ornstein-Zernicke approach [46],

$$\langle \Delta P(\mathbf{r}+\mathbf{s})\Delta P(\mathbf{s})\rangle = \frac{k_B T}{4\pi G s} \exp\left(-\frac{s}{\lambda(T)}\right), \qquad (A2)$$

with the Boltzmann constant $k_B$ and the above introduced characteristic length $\lambda(T) = \sqrt{G/|A|}$, which is, in contrast to $L(\tau)$, independent of time but temperature dependent. Its temperature dependence can be explicitly expressed as $\lambda(T) = \lambda_0 \sqrt{T_c/(T_c-T)}$ with the minimum correlation length reached at $T = 0$, $\lambda_0 = \sqrt{G/\alpha_0 T_c}$.

If the correlation length of the initial quenched disorder, $L(0)$, is much shorter than $\lambda_0$, then the quenched fluctuations, Eq. (A1), at the relevant distance $s \simeq \lambda(T)$ will be exponentially small and thus negligible. Therefore, we will consider an opposite limit of large-scale quenched disorder with $L(0) \gg \lambda_0$ which seems to be realized experimentally [3-9,14,15]. Prevailing of the quenched disorder over the thermodynamic fluctuations is then given by comparison of Eq. (A1) and (A2) taken at $s \simeq \lambda(T)$ resulting in inequality

$$D(0)P_s^2 \exp\left(-\frac{\lambda_0^2}{6L^2(0)}\frac{T_c}{T_c-T}\right) \gg \frac{k_B T}{4\pi G \lambda_0}\sqrt{\frac{T_c-T}{T_c}}. \qquad (A3)$$

This condition can hardly be satisfied by temperatures very close to $T_c$, that is why the validity of the model is in any case limited to the temperature range

$$\frac{T_c-T}{T_c} > \frac{\lambda_0^2}{6L^2(0)}. \qquad (A4)$$

Considering this constraint, the inequality can be further reduced to

$$\frac{T_c-T}{T_c} \gg \left(\frac{k_B}{8\pi\Delta C_v \lambda_0^3 D(0)}\right)^2 \quad (A5)$$

with the jump of the isochoric heat capacity at the second order phase transition $\Delta C_v = \alpha_0^2 T_C/(2B)$. With the maximum conceivable dimensionless dispersion of $D(0) \simeq 1$ the condition (A5) is nothing else but the Levanyuk criterion of applicability of the Landau theory [46], which fails in the close vicinity of $T_c$. The critical region near $T_c$ where the LGD approach fails in TGS was estimated in Refs. [47-49] as $|T_c - T| <0.02$ K. Combining Eqs. (A4) and (A5) the criterion for applicability of the model can be stated as

$$\frac{T_c-T}{T_c} \gg max\left\{\frac{\lambda_0^2}{6L^2(0)}, \left(\frac{k_B}{8\pi\Delta C_v \lambda_0^3 D(0)}\right)^2\right\}. \quad (A6)$$

## Appendix B. Examples of correlation functions for stripe domain structures

Correlation functions in section 3 are defined by statistical averaging of random variables indicated with a symbol $\langle…\rangle$. In experiment, however, they are typically derived from averaging of the observable sample area [4-10,16]. In the following, we will use this empirical definition of the correlations which can be, in particular, applied to regular domain structures. We consider a couple of examples to learn what spatial dependences of the correlation functions and what values of polarization dispersion may be expected.

We begin with a one-dimensional stripe domain structure described by a harmonic function,

$$\pi(x) = \cos(\pi x/2h) \quad (B1)$$

The period of this structure along the $x$-axis is $4h$ exhibiting positive and negative domains of the width $2h$. Let us assume a finite sample of the length $L = 4hN, N \gg 1$ with periodic boundary conditions. For this structure, obviously, $\bar{\pi} = 0$ and $\pi(x) = \xi(x)$. The correlation function can be introduced as

$$K(s) = \langle\xi(x+s)\xi(x)\rangle = \frac{1}{L}\int_{-L/2}^{L/2} dx \cos\left(\frac{\pi}{2}\frac{x+s}{h}\right)\cos\left(\frac{\pi x}{2h}\right)$$

$$= \frac{1}{2}\cos\left(\frac{\pi s}{2h}\right) + \frac{h}{\pi L}\left[\sin\left(\frac{\pi L+s}{4 h}\right) + \sin\left(\frac{\pi L-s}{4 h}\right)\right] \to \frac{1}{2}\cos\left(\frac{\pi s}{2h}\right), \ at \ L \to \infty. \quad (B2)$$

As expected for a regular structure, the correlation function appears to be periodic with the same period $4h$ exhibiting finite correlations at an infinite distance $s$. It should be noted that the domain wall width is of the same order as the structure period and the polarization is smaller than unity everywhere but extremum points. We note also a general relation for the dispersion

$$D = \langle\xi(x)\xi(x)\rangle = \langle\pi^2\rangle - \bar{\pi}^2, \quad (B3)$$

used below. For the considered structure D=1/2.

Let us consider now a stripe structure with alternative properties, a nonvanishing mean polarization $\bar{\pi}$ and domain walls of zero thickness. It is defined as

$$\pi(x) = \begin{cases} 1, & -a/2 < x < a/2 \\ -1, & -(a+b)/2 < x < -a/2 \\ -1, & a/2 < x < (a+b)/2 \end{cases}, \tag{B4}$$

where $a$ and $b$ indicate the width of the positive and negative domains, respectively. The average value of polarization is given by averaging over one period

$$\bar{\pi} = \frac{1}{a+b} \int_{-(a+b)/2}^{(a+b)/2} dx\, \pi(x) = \frac{a-b}{a+b}. \tag{B5}$$

The variance, or dispersion, of $\pi(x)$ is easy to calculate without integration, noticing that in this structure $\langle \pi^2 \rangle = 1$ and, thus,

$$D = 1 - \bar{\pi}^2 = \frac{4ab}{(a+b)^2}. \tag{B6}$$

It can be seen that it turns to zero in a single-domain state, when $a = 0$ or $b = 0$, and reaches a maximum of $D = 1$ when $a = b$.

This structure can be infinitely continued periodically with a period (a+b) as follows:

$$\pi(x) = \sum_{n=-\infty}^{\infty} \left\{ -\vartheta\left[x + \frac{a+b}{2} + n(a+b)\right] + 2\vartheta\left[x + \frac{a}{2} + n(a+b)\right] - 2\vartheta\left[x - \frac{a}{2} + n(a+b)\right] + \vartheta\left[x - \frac{a+b}{2} + n(a+b)\right] \right\}. \tag{B7}$$

Direct calculation of the correlation function for such a piecewise-defined function turns out to be cumbersome, but can be conveniently performed using the Fourier series representation,

$$\pi(x) = \sum_{n=-\infty}^{\infty} c_n\, e^{ik_n x} \quad \text{with} \quad k_n = 2\pi n/(a+b), \tag{B8}$$

where

$$c_n = \frac{1}{a+b} \int_{-(a+b)/2}^{(a+b)/2} dx\, \pi(x) \cos(k_n x). \tag{B9}$$

because the polarization is an even function. Substituting the expression (B4) into (B9), we get, after some calculations, a formula

$$c_n = -\delta_{n,0} + \frac{2}{\pi n} \sin\left(\pi n \frac{a}{a+b}\right). \tag{B10}$$

From this we can see, particularly, two cases of single-domain states, $\pi(x) = 1$ at $b = 0$ with $c_n = \delta_{n,0}$ and $\pi(x) = -1$ at $a = 0$ with $c_n = -\delta_{n,0}$. In general, $\bar{\pi} = c_0 = (a-b)/(a+b)$.

Now, the correlation function reads

$$K(s) = \frac{1}{a+b} \int_{-(a+b)/2}^{(a+b)/2} dx\, \xi(x+s)\xi(x) = \frac{1}{a+b} \int_{-(a+b)/2}^{(a+b)/2} dx\, \pi(x+s)\pi(x) - \bar{\pi}^2. \tag{B11}$$

Using the Fourier series (B8), Eq. (B11) is reduced to

$$K(s) = \sum_{n=-\infty}^{\infty} |c_n|^2 e^{ik_n s} - \bar{\pi}^2. \tag{B12}$$

which is transformed to the series

$$K(s) = \frac{4}{\pi^2} \sum_{n=1}^{\infty} \frac{1}{n^2} \left[ \cos\left(2\pi n \frac{s}{a+b}\right) - \frac{1}{2}\cos\left(2\pi n \frac{a-s}{a+b}\right) - \frac{1}{2}\cos\left(2\pi n \frac{a+s}{a+b}\right) \right]. \tag{B13}$$

Using the formula 5.4.2.7 from the tables [50] the summation in Eq. (B13) can be performed to give

$$K(s) = 4 \left[ B_2\left(\frac{s}{a+b}\right) - \frac{1}{2} B_2\left(\frac{a-s}{a+b}\right) - \frac{1}{2} B_2\left(\frac{a+s}{a+b}\right) \right], \tag{B14}$$

where $B_2(x) = (1/6) - x + x^2$ is the Bernulli polinomial. Using this, one should account that the argument in Eq. (B14) may vary only within the region $0 \leq x \leq 1$. To satisfy this requirement for $0 \leq s \leq a + b$, the argument in the last term of (B13) can be shifted by $a + b$ resulting finally in the expression

$$K(s) = 4 \left[ B_2\left(\frac{s}{a+b}\right) - \frac{1}{2} B_2\left(\frac{|a-s|}{a+b}\right) - \frac{1}{2} B_2\left(\frac{|b-s|}{a+b}\right) \right]. \tag{B15}$$

Taking for example $b \leq a$ the correlation function results as

$$K(s) = \frac{4}{(a+b)^2} \begin{cases} ab - s(a+b), & 0 \leq s < b < a \\ -b^2, & b \leq s < a \\ s(a+b) - ab - a^2 - b^2, & b < a \leq s \end{cases}. \tag{B16}$$

Thus, the correlation function for an unbalanced infinite stripe domain structure is also periodic one with the period $(a + b)$ and varies between $K(0) = 4ab/(a+b)^2$ and $K(b) = -4b^2/(a+b)^2$ changing sign at $s = ab/(a+b)$ and $s = (a+b) - ab/(a+b)$. In realistic random conditions, correlation function may oscillate but decays at a finite correlation length.

## Appendix C. Energy contribution due to polarization fluctuations

Let us evaluate the contribution of spatial polarization fluctuations to the energy using the functional (1) in the dimensionless form. By averaging over the ferroelectric volume, one finds

$$\Delta \Phi = P_s^2 |A| \int_{V_f} \left[ -\frac{1}{2} D(\tau) + \frac{1}{4}\left(6\bar{\pi}^2(\tau) + 3D(\tau)\right) D(\tau) + \frac{1}{2L^2(\tau)} D(\tau) - \Psi_{zz}(\tau) \right] dV. \tag{C1}$$

Considering that asymptotically $L(\tau \to \infty) \to \infty$ and $\Psi_{zz}(\tau \to \infty) \to 0$ and using Eqs. (50), one obtains the fluctuation contribution to the final state energy

$$\Delta \Phi = -\frac{3}{4} P_s^2 |A| D^2(\tau \to \infty) V_f \tag{C2}$$

exhibiting the tendency in favor of the multi-domain state in comparison with the single-domain state with $D = 0$.

## Acknowledgements

We thank Prof. Dragan Damjanovic for useful discussion of the results. O.Y.M. is grateful for the financial support by DAAD during the visit at the TU Darmstadt. This work was supported by the Deutsche Forschungsgemeinschaft (German Research Society, DFG) via the grant No. 405631895 (GE-1171/8-1).